\documentclass[lettersize,journal]{IEEEtran}
\usepackage{amsmath,amsfonts}
\usepackage{algorithmic}
\usepackage{algorithm}
\usepackage{array}
\usepackage{textcomp}
\usepackage{stfloats}
\usepackage{url}
\usepackage{verbatim}
\usepackage{graphicx}
\usepackage{cite}
\usepackage{booktabs}
\usepackage{footnote}
\usepackage[bottom]{footmisc}
\usepackage{hyperref}
\usepackage{xcolor}
\usepackage{graphicx}
\usepackage{subfigure}
\usepackage{enumitem}
\usepackage{subcaption}

\begin{document}
\title{Generative AI on the Edge: Architecture and Performance Evaluation}

\author{
    \IEEEauthorblockN{Zeinab Nezami\IEEEauthorrefmark{1}, Maryam Hafeez\IEEEauthorrefmark{1}, Karim Djemame\IEEEauthorrefmark{2}, Syed Ali Raza Zaidi\IEEEauthorrefmark{1}}\\
    \{\IEEEauthorblockA{\IEEEauthorrefmark{1}School of Electronic and Electrical Engineering,
    \IEEEauthorblockA{\IEEEauthorrefmark{2}School of Computing\}, University of Leeds, UK\\
    \{Z.Nezami, M.Hafeez, K.Djemame, S.A.Zaidi\}@leeds.ac.uk}
}
\thanks{This work is funded by the Communications Hub for Empowering Distributed Cloud Computing Applications and Research (CHEDDAR), supported by the EPSRC through UKRI’s Technology Missions Fund.}
\thanks{Manuscript received April 19, 2021; revised August 16, 2021.}}


\IEEEpubid{0000--0000/00\$00.00~\copyright~2021 IEEE}

\maketitle

\begin{abstract}

6G's AI native vision of embedding advance intelligence in the network while bringing it closer to the user requires a systematic evaluation of Generative AI (GenAI) models on edge devices. Rapidly emerging solutions based on Open RAN (ORAN) and Network-in-a-Box strongly advocate the use of low-cost, off-the-shelf components for simpler and efficient deployment, e.g., in provisioning rural connectivity. In this context, conceptual architecture, hardware testbeds and precise performance quantification of Large Language Models (LLMs) on off-the-shelf edge devices remains largely unexplored. This research investigates computationally demanding LLM inference on a single commodity Raspberry Pi serving as an edge testbed for ORAN. We investigate various LLMs, including small, medium and large models, on a Raspberry Pi 5 Cluster using a lightweight Kubernetes distribution (K3s) with modular prompting implementation. We study its feasibility and limitations by analyzing throughput, latency, accuracy and efficiency. Our findings indicate that CPU-only deployment of lightweight models, such as Yi, Phi, and Llama3, can effectively support edge applications, achieving a generation throughput of 5 to 12 tokens per second with less than 50\% CPU and RAM usage. We conclude that GenAI on the edge offers localized inference in remote or bandwidth-constrained environments in 6G networks without reliance on cloud infrastructure.

\end{abstract}

\renewcommand\baselinestretch{0.9}

\section{Introduction}
\IEEEPARstart {R}{esearch} and development towards 6th generation of communication networks (6G) heavily emphasize the end-to-end integration of AI. Many AI capabilities are part of standardization efforts, building on existing foundations such as intelligent radio resource management in radio access network (RAN) and Network Data Analytics Function (NWDAF) in the core~\cite{5gstandard}. Within RAN, this integration of AI is accelerated by Open RAN (ORAN) which leverages disaggregation of the traditional architecture into modular interoperable components\footnote{O-RAN Alliance, \url{https://www.o-ran.org/}, accessed Nov 2024}. It further introduces the concept of extended applications xApps and rApps within the RAN Intelligence Controller to automate various management and optimization tasks via leveraging AI and machine learning.

\par AI has evolved from convolutional~\cite{krizhevsky2017imagenet} and recurrent neural networks~\cite{gers2000learning} to transformers~\cite{vaswani2017attention}, enabling greater scalability and the development of foundation models. These models, such as LLMs, use large internet datasets to perform diverse tasks~\cite{Laskaridis2024MELTingPM} without specific training for each.
The computation of AI algorithms is shifting from centralized data centers to decentralized edge devices, enhancing personalization and robustness. Gartner predicts that by 2025, over half of deep neural network data analysis will occur on edge devices, or Edge AI, closer to data sources~\cite{GartnerHypeCycleforEdgeComputing2024}. This shift, particularly in 6G networks, paves the way for LLMs on edge devices, democratizing AI and enabling real-time applications with 6G’s ultra-low latency, high bandwidth, and edge computing. This will improve services such as NLP, real-time translation, personalized assistance, and optimize network performance through dynamic traffic prediction and resource management~\cite{kan2024mobile}.

\par While AI-native design of future networks with extensive use of GenAI is inevitable, it is becoming increasingly crucial to quantify the performance of these models on resource-constrained edge devices that lack a GPU. Significant intelligence is bundled in ORAN for low latency and enhanced service provision. However, it is not practical to have a GPU in every deployment scenario. For instance, Vodafone has recently showcased its low cost private 5G solutions running on a Raspberry Pi\footnote{Vodafone, \url{https://www.vodafone.com/news/technology/vodafone-unveils-prototype-5g-network-built-raspberry-pi-computer}, accessed Nov 2024}. These low cost platforms are ideal and will become increasingly common for extending connectivity in an energy-efficient manner for the benefit of all sectors of modern economies including rural deployments. However, a systematic evaluation of provisioning LLM capabilities on such platforms remains an open question. LLM inference demands high computational and memory resources, typically requiring multiple high-end accelerators, which pushes the limits of the infrastructure~\cite{kwon2023efficient}. While advancements in lightweight models, optimization techniques, and quantization have made it possible to run smaller versions of LLMs on low-cost, low-power devices without entirely relying on cloud infrastructure, the performance of these advancements and their relative benefit in edge deployments are not yet fully understood. 

\par In this context, Edge AI testbeds are crucial for validating the feasibility of real-time, distributed intelligence within next-generation networks. In response to the growing demand for privacy-sensitive real-time AI applications, this research investigates the deployment of LLMs in edge environments, specifically for PromptAI (a conversational AI assistant) on low-cost edge infrastructure. We examine LLM deployment on Raspberry Pi using a lightweight Kubernetes distribution (K3s), demonstrating both the feasibility and limitations of such setups, and providing insights to guide future research and optimizations for LLMs in resource-constrained edge infrastructures. This work aligns with the vision of 6G ORAN by demonstrating how AI models can be effectively deployed to facilitate real-time decision-making and ultra-low latency communication at the network edges\footnote{6G SNS IA whitepaper,  \url{https://6g-ia.eu/plans-papers/}, accessed Oct 2024}.

\par The main contributions of this paper are as follows:
(i) an edge computing testbed for LLM experimentation using Raspberry Pis orchestrated by K3s,
(ii) a modular PromptAI framework for LLM application experimentation, built with an API-based architecture for seamless extensibility.
(iii) preliminary evaluations of popular LLMs of different sizes, providing performance quantification and comparison.
(iv) an open-source GitHub repository\url{https://github.com/cheddarhub/edgeAI}, a DockerHub repository for Docker images\url{https://hub.docker.com/repositories/edgeAI}, facilitating further research in GenAI. The paper is structured as follows: Section 2 reviews related work and identifies the research gap that this paper addresses. Section 3 describes the system architecture, outlining the design choices and configurations employed in our GenAI deployment. Section 4 details the experimental setup and evaluation results. Section 5 concludes the paper with key findings and future research directions.

\section{Related work}
\noindent Recent advances in LLMs have made LLM inference a significant workload, driving active research in the field. On-device DNN benchmarking has been explored extensively for mobile and edge deployments. Early works~\cite{ignatov2018ai} introduced benchmarking suite for device ranking and analyzing model execution across various downstream tasks, while MLPerf~\cite{reddi2020mlperf} has become a standardized industry benchmark. Recent studies~\cite{xu2019first} have also focused on evaluating performance in real-world mobile applications, highlighting the increasing adoption of on-device machine learning. However, the growing computational demands of LLMs have led most deployments to offload inference to the cloud~\cite{mao2017survey}, with on-device deployment remaining limited. This challenge is compounded by the lack of suitable tools, highlighting the need for improved on-device measurement to support edge-based LLM execution.
\par In the LLM domain, previous research has primarily addressed training efficiency~\cite{rajbhandari2020zero,you2023zeus} or optimized inference~\cite{kwon2023efficient,aminabadi2022deepspeed} within data center environments.
There has been growing interest in enabling edge execution of LLMs through various frameworks. llama.cpp\footnote{llama.cpp, \url{https://github.com/ggerganov/llama.cpp}, accessed Oct 2024} and MLC\footnote{MLC-LLM, \url{https://github.com/mlc-ai/mlc-llm}, accessed Oct 2024} offer cross-platform accelerated execution with support for diverse LLM architectures. Other frameworks such as llama2.c\footnote{llama2.c, \url{https://github.com/karpathy/llama2.c}, accessed Oct 2024} focus on simplicity, while tinygrad\footnote{Tinygrad, \url{https://github.com/tinygrad/tinygrad}, accessed Oct 2024} aims at accelerated execution but lacks quantized model support. TinyChatEngine\footnote{TinyChatEngine, \url{https://github.com/mit-hanlab/TinyChatEngine}, accessed Oct 2024} provides on-device inference with compressed models. MELT~\cite{Laskaridis2024MELTingPM} introduces an automation infrastructure developed for executing and benchmarking LLMs on personal mobile devices. Recently, operating system providers have introduced their own platforms, such as Google’s AICore\footnote{AICore, \url{https://developer.android.com/ml/aicore}, accessed Oct 2024}, though these are limited to specific devices. Additionally, Google has released MediaPipe\footnote{MediaPipe, \url{https:
//developers.googleblog.com/2024/03/running-large-languagemodels-on-device-with-mediapipe-andtensorflow-lite.html}, accessed Oct 2024} to support on-device LLM execution. 

\par Related work reveals that the inference of LLMs is primarily constrained by memory and CPU power~\cite{Laskaridis2024MELTingPM,kwon2023efficient,aminabadi2022deepspeed}, indicating that the field is still evolving. While GPUs are often considered the optimal choice for running complex machine learning models due to their superior processing power, their high cost can be prohibitive for many applications. In contrast, spare edge devices equipped with CPUs are widely available and affordable, making them an attractive option for enabling broader access to AI capabilities in edge computing scenarios. This work presents a novel approach by establishing a K3s-based Raspberry Pi testbed, specifically designed to deploy and evaluate LLMs on edge devices operating solely on CPUs. Utilizing emerging LLMs, open-source tools, and a lightweight Kubernetes environment, our setup measures the resource consumption and inference speed of LLMs in real-world edge scenarios. Furthermore, this research provides a detailed performance analysis of LLMs on resource-constrained edge devices—an aspect that current studies on LLMs do not comprehensively address.

\section{System architecture}
\begin{figure}[!t]
\centering
\includegraphics[clip, trim=7.5cm 1.4cm 7.7cm 2.6cm, width=\columnwidth]{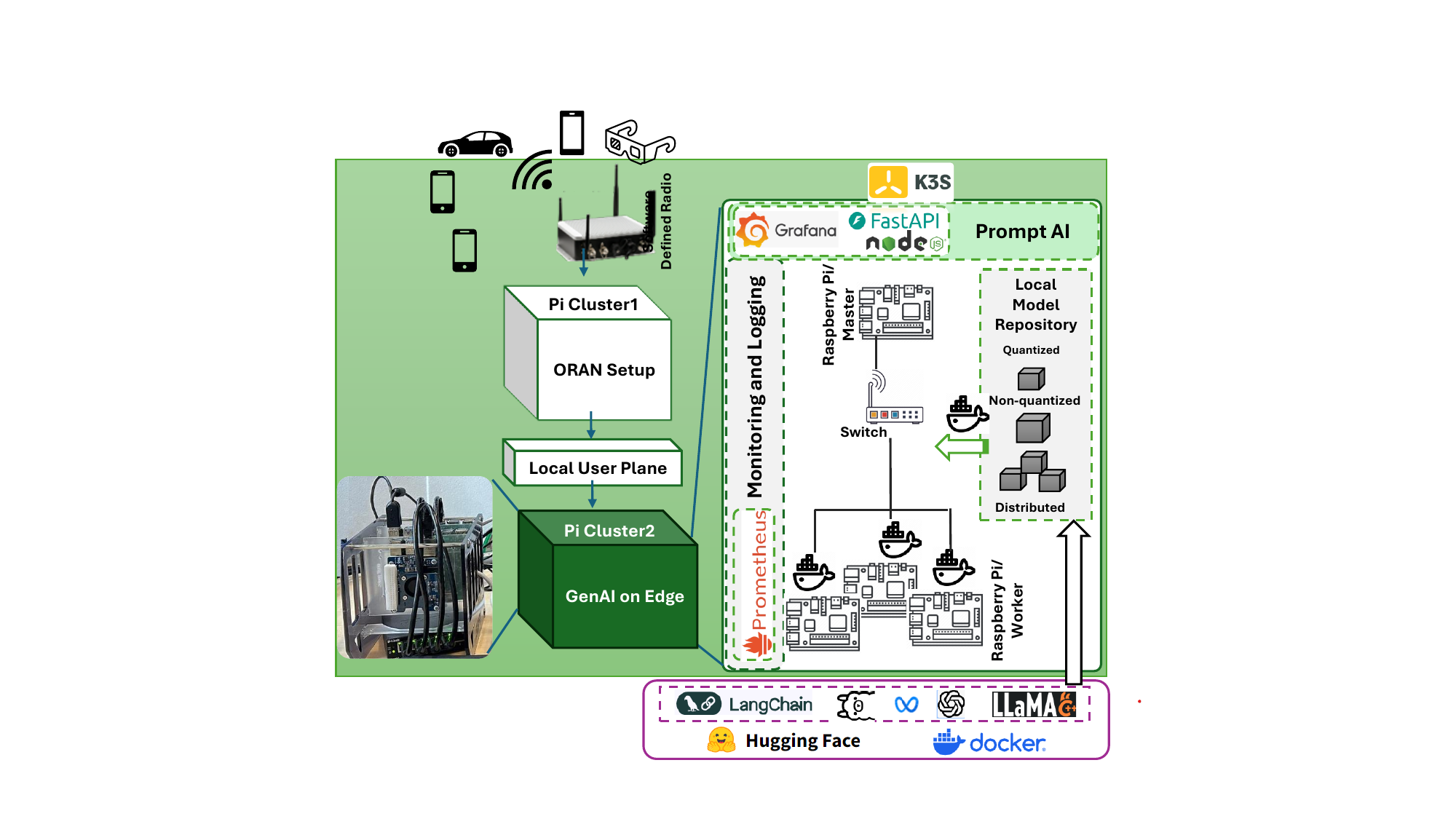}
\caption{Architecture of the GenAI testbed, its key components and their interconnections in 6G networks}
\label{fig:testbed}
\vspace{-15pt}
\end{figure}

\noindent We consider a small base station (SBS) which is composed of two compute clusters. Considering the option for low-cost deployment, we consider these compute clusters to be Raspberry Pi single board computers (SBC). We do not assume any graphical processing unit (GPU) capability across these clusters. The first cluster hosts containerised implementation of ORAN. A reference implementation can run SRS RAN stack as demonstrated by Vodafone in recent trials. The second cluster operates to serve U-plane traffic and performs edge offloading for LLM based tasks. Our focus in this article is on performance of LLMs deployed on this cluster and therefore rest of the discussion is geared towards this. The cluster forms API-based edge computing testbed for LLM deployment on constrained hardware, consisting of four Raspberry Pis functioning as edge nodes, including an orchestrator (K3s master). The system utilizes Docker containers to deliver GenAI services in isolated environments across the Raspberry Pis, managed by K3s. 
Figure.~\ref{fig:testbed} presents the overall system architecture and the core components of the container-based PromptAI running on the testbed. 
The testbed prototype is implemented using a cluster of Raspberry Pi 5 units, each equipped with a Quad-core ARM Cortex-A76 processor operating at 2.0 GHz, with 8 GB of RAM and 128 GB of storage. A D-Link 5-Port Desktop Gigabit PoE+ Switch used to connect the Raspberry Pis into a cluster. The edge nodes are not limited to Raspberry Pi devices; any compatible system can join the K3s network, supporting a scalable and heterogeneous architecture. Integrated with sensors, Pis enable diverse applications, including Edge AI. Deploying LLM services as Docker containers ensures secure, isolated execution on resource-constrained devices, improving flexibility and efficiency in edge hardware deployment.
\noindent \textbf{K3s}: K3s is a streamlined version of Kubernetes specifically designed for resource-constrained IoT and edge computing environments. It boasts a significantly reduced memory footprint, fast startup times, and simplified configuration, making it ideal for lightweight, distributed deployments. The K3s master node in this testbed acts as the central orchestrator, interfacing with the system’s API to manage the allocation and release of edge resources. K3s efficiently orchestrates services across edge nodes, ensuring smooth operation in resource-constrained environments, with automated updates and easy scalability.\\
\noindent \textbf{Docker container}: The testbed deploys a distributed PromptAI system using Docker containers, ensuring fast initialization and lightweight operation across the Raspberry Pi cluster. This system includes interconnected services running via APIs, including a prompt front-end, proxy, and model containers. The PromptAI front-end, developed using FastAPI and Node.js, allows users to select from a pool of LLMs sourced from Hugging Face and stored in a Docker Hub repository, enabling them to engage with their chosen model. Requests are forwarded through a proxy service that directs traffic to the appropriate model container on the Raspberry Pi nodes. By leveraging Docker containers, each component runs in an isolated, reproducible environment, providing scalability, flexibility, and ease of management—qualities essential for efficient edge computing in resource-constrained environments.\\
\noindent \textbf{Grafana/Prometheus}: The testbed is equipped with open-source tools such as Prometheus and Grafana to facilitate comprehensive monitoring and visualization of system metrics. Prometheus collects and stores time-series data for tracking key performance metrics, while Grafana offers interactive data visualization. Together, they enable real-time monitoring of CPU, memory, and network usage across edge resources, with customizable Grafana dashboards for tailored GenAI testbed visualization.\\
\noindent \textbf{Models}: This work utilizes quantized models that use less memory and computational power, making them more suitable for running on resource-constrained devices, while maintaining a balance between performance and resource use. We focus on the GPT-Generated Unified Format (GGUF), a more recent and advanced quantization method for LLMs\footnote{Ggml library: Large language models for everyone, \url{https://github.com/rustformers/llm/tree/main/crates/ggml}, accessed Oct 2024}, using its optimized storage and quantization to enable faster LLM inference. We select 4-bit quantized models, which strike a balance between minimal accuracy loss and significant size reduction.
As outlined in Table~\ref{tab:models}, we selected a diverse set of decoder-only models from the open\_llm\_leaderboard\footnote{\url{https://huggingface.co/spaces/open-llm-leaderboard/open_llm_leaderboard}} on the Huggingface website, along with the latest popular, regularly updated models. These models are then deployed and evaluated in three groups, categorized by size: large, medium, and small. Large models include \textbf{InternLM}, developed by Shanghai AI Lab and SenseTime, which excels in multilingual tasks such as comprehension, reasoning, and coding; \textbf{Mistral}, optimized for high-performance text generation; and Meta’s \textbf{Llama2}, focused on inference and fine-tuning tasks. Medium-sized models feature Microsoft’s \textbf{Phi}, an ONNX-based model efficient across diverse hardware and ideal for resource-limited settings, along with Meta’s \textbf{Llama3}, a compact multilingual model for dialogue and summarization that balances capability with efficiency. Small models include \textbf{Yi}, which targets code generation and natural language tasks with minimal resource usage; Stability AI’s StableLM \textbf{Zephyr}, optimized for interactive applications on mobile and edge devices; and \textbf{Gemma}, a smaller LLaMA-based model that performs well in text completion.
\begin{table}[t]
    \centering
    \caption{Overview of Tested Models}
    \label{tab:models}
    \footnotesize
    \begin{tabular}{p{1.2cm}p{0.5cm} p{4.9cm}}
        \toprule
        \multicolumn{1}{l}{\textbf{Model}} & \multicolumn{1}{l}{\textbf{Size}} & \multicolumn{1}{l}{\textbf{HF Repository}} \\
        \midrule
        \multicolumn{3}{c}{\textbf{Large Models (Above 6B)}} \\ 
        \midrule
        InternLM & 7.74B & second-state/internlm2\_5-7b-chat-GGUF \\
        Mistral & 7.25B & bartowski/Mistral-7B-Instruct-v0.3-GGUF \\
        Llama2 & 6.74B & TheBloke/Llama-2-7b-Chat-GGUF \\
        \midrule
        \multicolumn{3}{c}{\textbf{Medium Models (Between 3B and 6B)}} \\ 
        \midrule
        Phi& 3.82B & bartowski/Phi-3.5-mini-instruct-GGUF \\
        Llama3 & 3.21B & bartowski/Llama-3.2-3B-Instruct-GGUF \\
        \midrule
        \multicolumn{3}{c}{\textbf{Small Models (Below 3B)}} \\ 
        \midrule
        Zephyr& 2.8B & TheBloke/stablelm-zephyr-3b-GGUF \\
        Gemma & 2.61B & bartowski/gemma-2-2b-it-GGUF \\
        Yi &1.48B & bartowski/Yi-Coder-1.5B-Chat-GGUF \\
        \bottomrule
    \end{tabular}
    \vspace{-10pt}
\end{table}
\section{Evaluation}\label{evaluation}
\begin{figure}[t]
\centering
\includegraphics[clip, trim=0.5cm 1.5cm 0cm 1.2cm, width=\columnwidth]{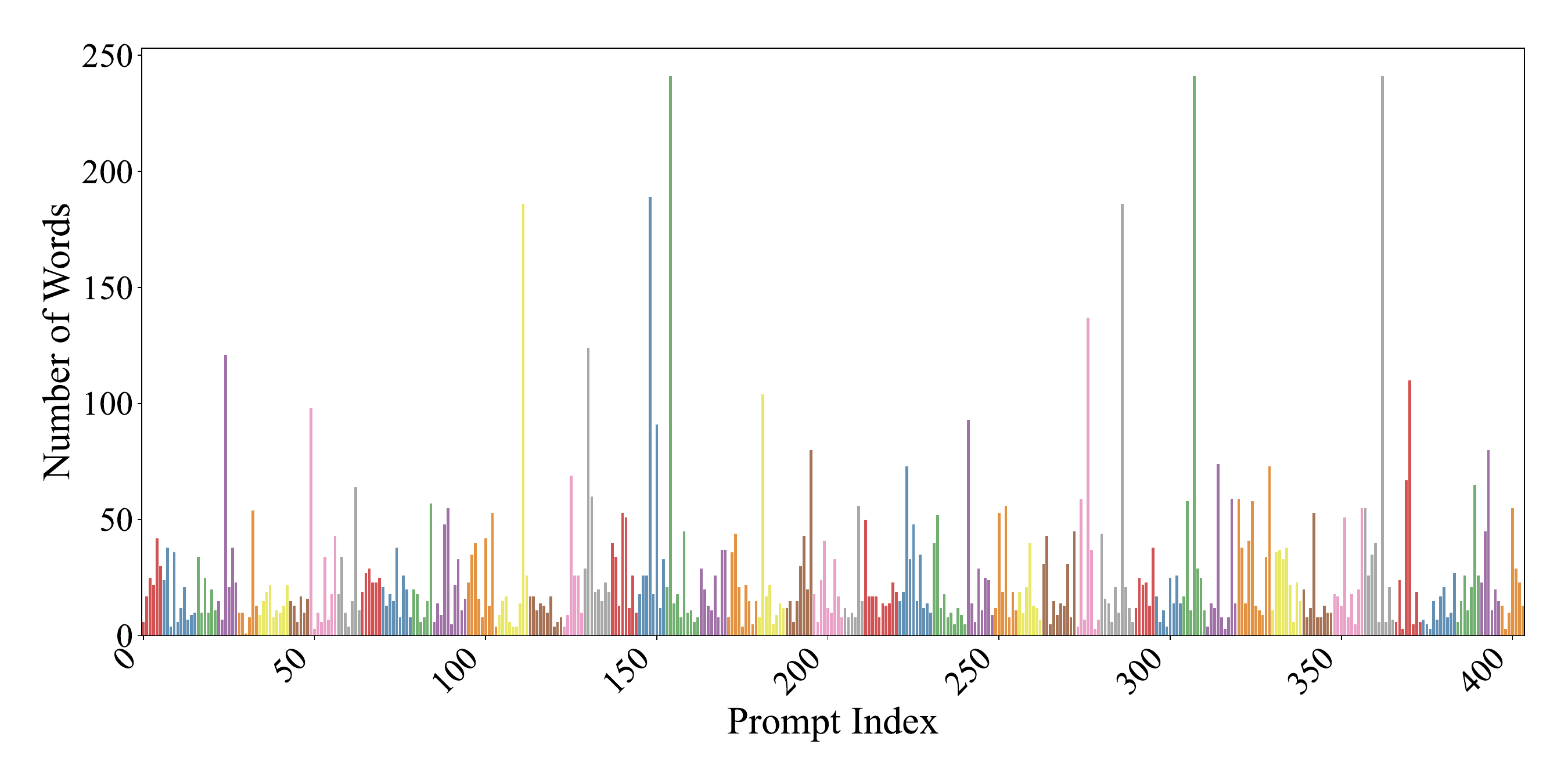}
\caption{Prompt lengths across conversations (highlighted by different colors)}
\label{fig:conv-prompt}
\vspace{-10pt}
\end{figure}

This section examines LLM inference performance of popular LLM models using real conversations.
We employ a subset of prompts from the OpenAssistant/oasst1 dataset\footnote{OpenAssistant dataset,  \url{https://huggingface.co/datasets/OpenAssistant/oasst1}, accessed Oct 2024}~\cite{Laskaridis2024MELTingPM}, consisting of 50 distinct dialogues, each with a minimum of five interaction turns per conversation. 
As shown in Figure.~\ref{fig:conv-prompt}, the dataset has a moderate average prompt length of 25 words but exhibits high variability, with a standard deviation of 30.67 words and prompt lengths ranging from 1 to 241 words. Each experiment is repeated three times to ensure consistency and reliability of results.
We assess five key metrics—memory usage, CPU utilization, latency, accuracy, and computational throughput—offering a comprehensive view of resource demands and efficiency for PromptAI in resource-constrained settings. Since GGUF model evaluation is in its early stages, with limited supporting libraries, we use the llama.cpp framework to capture fine-grained data for PromptAI operations during inference. To ensure deterministic execution, maximum token generation is set to 500.
\subsection{Throughput and Latency}

\begin{figure}[t]
\centering
    \includegraphics[clip, trim=0cm 0.8cm 0cm 0.5cm, width=\columnwidth]{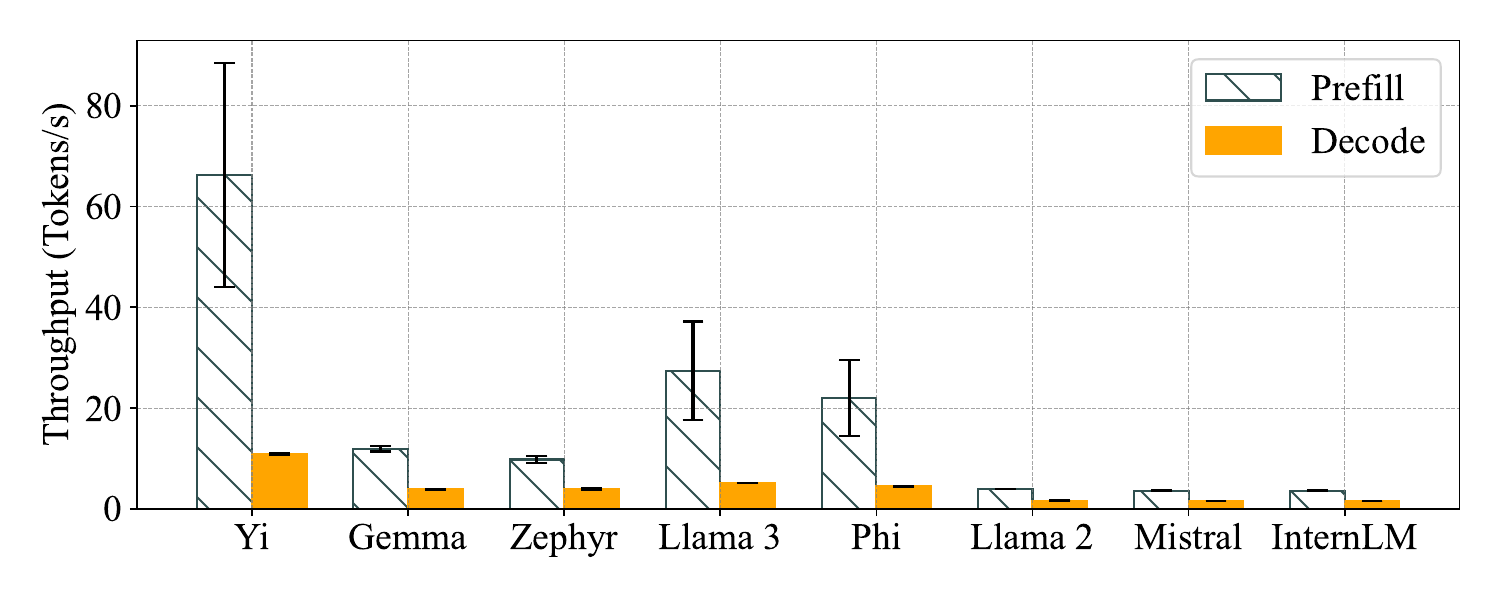}
\caption{Throughput performance for Prefill and Decode stages across models}
\label{fig:throughput}
\vspace{-10pt}
\end{figure}

Inference in LLMs involves two key phases: Prefill and Decode (i.e., generation). The Prefill phase computes the key-value cache for the prompt input tokens before autoregressive generation begins. The Decode phase then selects the next token(s) from a probability distribution generated by the model and converts them into human-readable output. Figure~\ref{fig:throughput} illustrates average throughput for Prefill and Decode phases, highlighting model efficiency in processing conversational prompts.
Among large models, Llama2 and InternLM show stable Prefill throughput with low variability. Llama2 achieves around 7\% higher Decode throughput than InternLM, though InternLM’s higher Prefill variability may introduce output speed fluctuations in time-sensitive applications. Mistral’s Prefill throughput aligns closely with InternLM’s but has 5.8\% higher Decode throughput.

\par For medium-size models, Llama3 significantly outperforms Phi in Prefill throughput by 25\%, suggesting it can handle initial token processing more efficiently and reduce initial latency. However, Llama3’s Prefill variability may affect consistency on constrained devices. Phi’s Decode throughput is strong, though Llama3 exceeds it by 13\%, providing advantages in both initial and ongoing processing.
In small models, Yi achieves standout Prefill throughput, roughly 4.5 times higher than Gemma and Zephyr. Yi also leads in Decode throughput with 170\% greater efficiency than other small models, though its high Prefill variability may impact response-time predictability. Gemma outperforms Zephyr in Prefill throughput by about 21\%, with similar Decode throughput between them, making both suitable for edge applications needing moderate throughput and stability.
In summary, Llama3 suits tasks prioritizing high throughput, despite some Prefill variability. Yi delivers high performance but may introduce inconsistency in real-time settings, while Gemma and Zephyr offer balanced throughput and stability, fitting for moderately demanding edge applications.
Furthermore, the throughput of LLMs is fundamentally constrained by the Decode phase; however, initiatives like Distributed Llama, currently under development\footnote{Distributed Llama, \url{https://github.com/b4rtaz/distributed-llama}, accessed Nov 2024}, leverage tensor parallelism across multiple devices and show promising potential to address this limitation.

\begin{figure*}[!htb]
\centering
    \includegraphics[clip, trim=2cm 3.7cm 0cm 0cm, width=21cm]{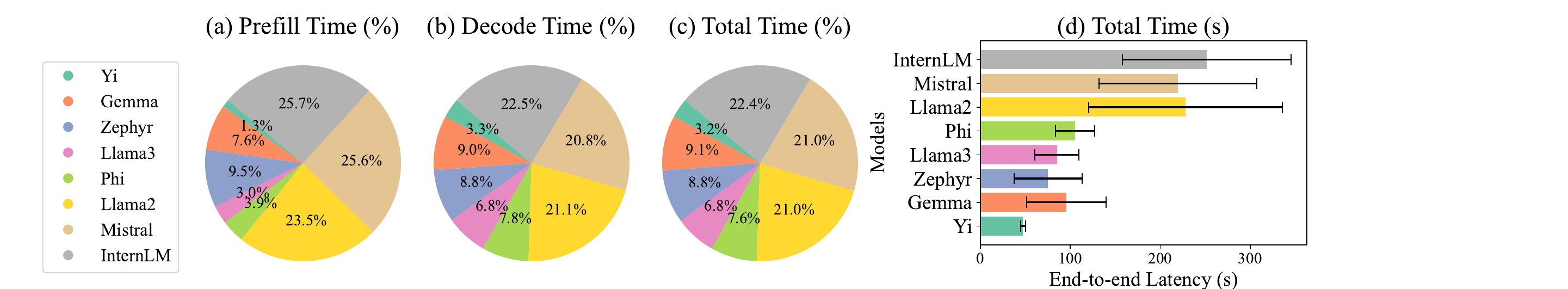}
\caption{Timing metrics across models: (a)-(c) Prefill, Decode, and total time per token for each model; (d) Total time (in seconds)}
\label{fig:latency}
\vspace{-10pt}
\end{figure*}
\par Figure~\ref{fig:latency}~(a-c) shows the percentage of time each model spends in the Prefill and Decode phases relative to total time. We normalized the timing data to calculate the average time per token, with Total Time representing the cumulative duration from input to output (end-to-end latency).
The latency analysis reveals notable performance differences among the models. In the large model category, Llama2 has a slightly faster Prefill time (252.64 ms per token), about 8\% faster than Mistral and InternLM. In contrast, Llama3 and Phi have much faster Prefill times of 32.59 ms and 41.50 ms, respectively, being 88\% and 85\% faster than InternLM, making them ideal for scenarios requiring efficient processing.
\par In the small model category, Gemma and Zephyr show higher latencies, with Prefill times of 82.02 ms and 102.62 ms, respectively. Their Decode times further increase total processing time, with Gemma taking 238.93 ms and Zephyr taking 233.88 ms, resulting in total times of 251.98 ms and 243.63 ms. In comparison, Yi achieves a total processing time over 80\% faster than both, with a Prefill time of just 13.79 ms—about 20 times faster than InternLM's 276.42 ms.
When considering end-to-end latency (Average Total Time), Figure~\ref{fig:latency}~d, the differences are more pronounced. Yi's 47.41 s is much faster than InternLM (251.96 s) and Llama2 (228.14 s), highlighting that smaller models such as Yi and Llama3 are competitive for applications requiring both low Prefill and overall low latency.
\subsection{Resource Consumption}
\begin{figure}[!t]
\centering
    \includegraphics[clip, trim=1.0cm 0cm 1cm 2cm, width=\columnwidth]{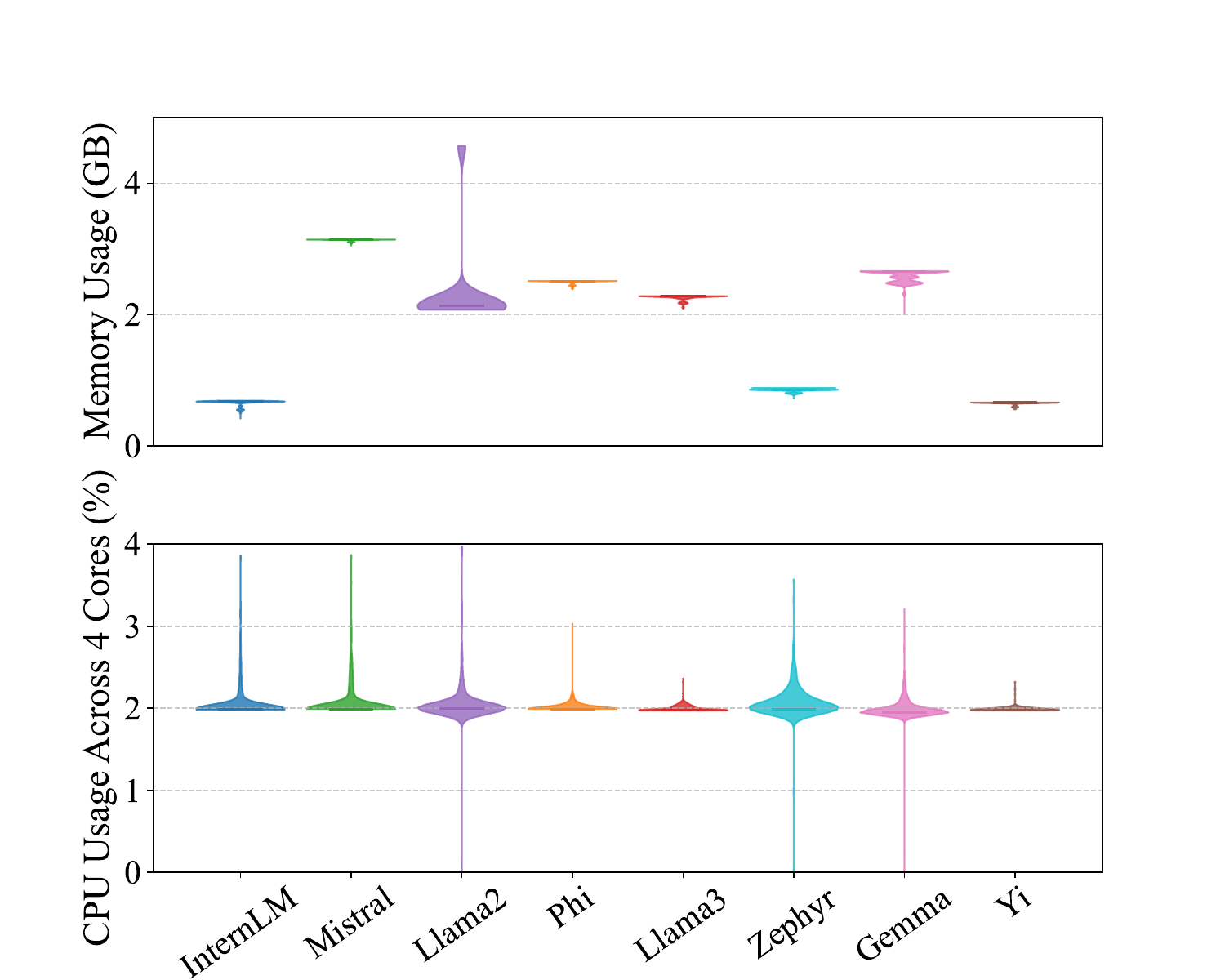}
\caption{Memory and CPU usage across different models}
\label{fig:rescons}
\vspace{-10pt}
\end{figure}

Figure.~\ref{fig:rescons} presents a comparative analysis of CPU and memory usage across various models during inference. Memory usage exhibits variability across model groups, with larger models typically demanding more resources. InternLM, Mistral, and Llama2 show higher memory usage, with values ranging from approximately 0.66 GB to 3.14 GB. In comparison, small models demonstrate lower memory requirements, with averages ranging from 0.65 GB to 2.59 GB. Notably, Yi has the lowest average memory usage among the small models at 0.65 GB, which is about 79\% lower than Mistral's 3.14 GB. Across all models, CPU utilization remains consistently low, with mean CPU usage centering around 50\%. There is minimal variation across the models in terms of CPU demand, suggesting low computational overhead despite the differences in model size. The results indicate that smaller models, with their lower memory and CPU demands, are well-suited for memory-constrained edge devices, while larger models may require more powerful hardware or optimizations using model pruning. Larger models may occasionally cause slight CPU spikes, but their overall CPU demand remains low, making them feasible for deployment in edge environments with limited processing resources.
\subsection{Deployment Stability and Accuracy}

The models studied in this research did not encounter out-of-memory (OOM) errors or hardware restarts during cluster deployment, demonstrating robustness and stability in handling large workloads. While quantization effectively reduces memory usage and accelerates inference, it often sacrifices accuracy, especially at lower precision levels (e.g., 4 bits). To evaluate accuracy, we use the Winogrande benchmark dataset\footnote{\url{https://huggingface.co/datasets/automated-research-group/winogrande}, accessed Nov 2024}, designed for common-sense reasoning, and the LM-Evaluation Harness~\cite{gao2021framework} framework with its inference server to serve our quantized models. The dataset includes Natural Language Inference (NLI) tasks, where models select the answer with the highest log likelihood, which is then compared to the correct label. Model accuracy on Winogrande benchmark varies significantly, with InternLM leading at 0.8, followed by Gemma at 0.7, and Llama3 at 0.69. Mistral, Llama2, and Phi all share an accuracy of 0.68. Yi and Zephyr score lower at 0.49 and 0.46, respectively. These differences largely stem from model architecture and size~\cite{Laskaridis2024MELTingPM}. It is important to note that we evaluate the non-fine-tuned variants of the models, which typically serve as a proxy for the accuracy degradation in downstream models~\cite{Laskaridis2024MELTingPM}. We selected pre-trained models rather than fine-tuned versions due to the alignment in fine-tuning processes, which often leads to performance enhancements that are not directly comparable to the base models.

\subsection{Contextual Sensitivity}
\begin{figure}[t]
\centering
   \includegraphics[clip, trim=1.2cm 1.9cm 0cm 0cm, width=\columnwidth]{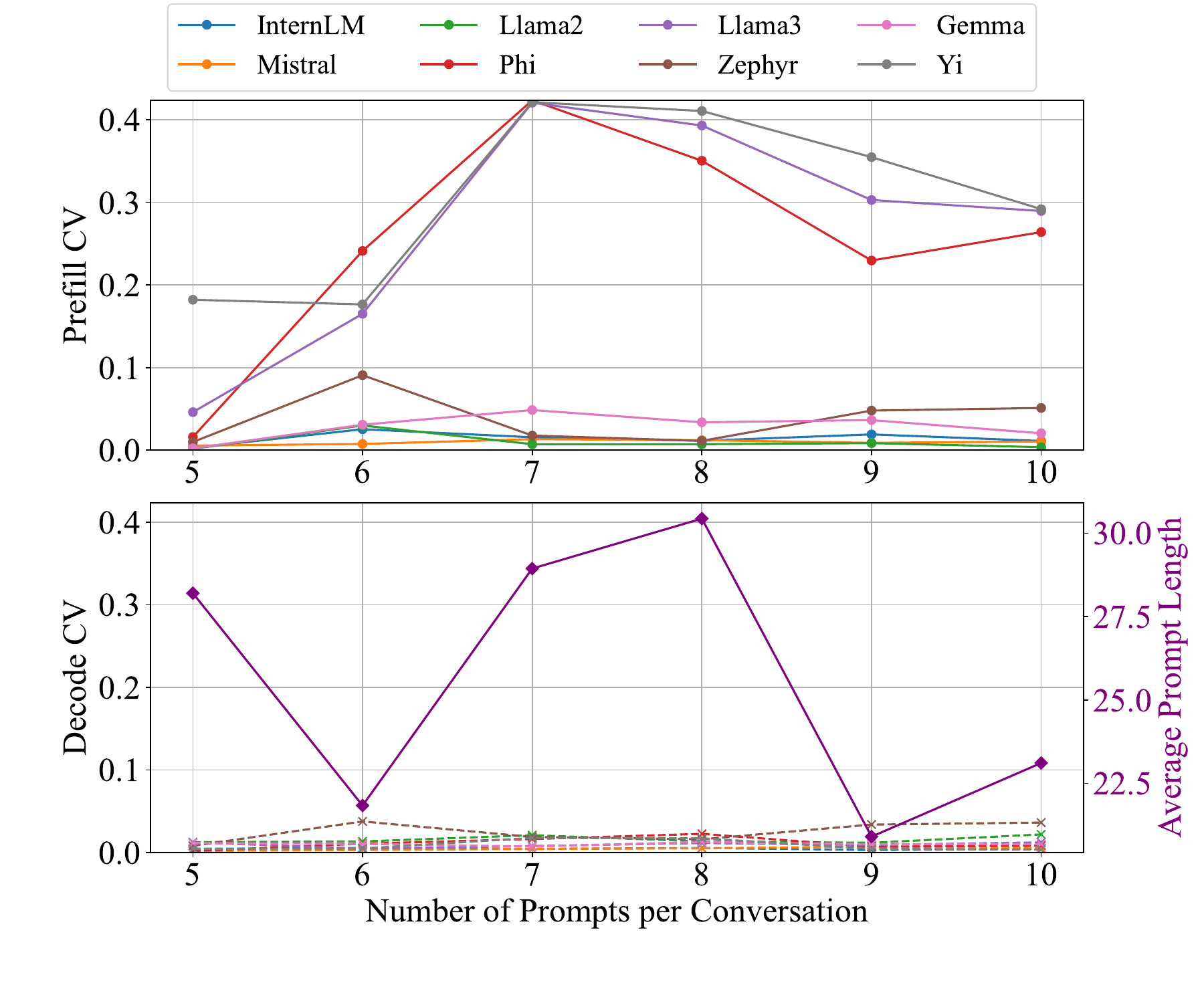}
\caption{Coefficient of Variance for Prefill and Decode throughput across LLMs}
\label{fig:context-thr}
\vspace{-10pt}
\end{figure}

This experiment examines how context length impacts language models throughput in conversational settings, such as dialogue agents and customer support. Figure.~\ref{fig:context-thr} measures throughput variability across models using the Coefficient of Variation (CV), the ratio of standard deviation to mean throughput, providing insights into stability across different context lengths. The Prefill throughput analysis reveals distinct clusters of performance consistency among the various models. In Llama3, Phi, and Yi, Prefill CV aligns with the prompt length fluctuations, indicating that longer prompts add variability to the model’s processing efficiency. Conversely, models such as Llama2, Mistral, Gemma, and Zephyr show stable CV values across various conversation lengths, demonstrating consistency in Prefill stages regardless of context size, making them more suited for contexts with varied conversation lengths.
On the other hand, Decode CV varies slightly in response to increasing conversation length and prompt size. InternLM and Mistral maintain the lowest Decode CVs, averaging around 0.0046, indicating very stable performance with minimal fluctuations, even as context length increases. Llama2 shows a higher CV (0.0157), indicating a more pronounced response to varying conversation lengths. Phi, Llama3, and Gemma show moderate variability, with Decode CVs ranging from approximately 0.0097 to 0.0112, suggesting these models are almost consistent, but with slight fluctuations as prompt lengths increase. Yi also exhibits moderate variability, similar to Phi and Llama3, although with slightly less fluctuation. Among the models, Zephyr shows the highest variability, indicating more substantial fluctuations in performance as prompt lengths increase.
Overall, Mistral and InternLM show stable throughput across varying context lengths, while Llama2 and Zephyr exhibit more variability, especially in the Decode stage. Llama3, Phi, and Yi maintain moderate performance consistency with slight fluctuations.

\subsection{Putting it All Together}
\begin{figure}[t]
\centering
\includegraphics[clip, trim=1.5cm 0.8cm 4cm 0cm, width=18cm, height=5.5cm]{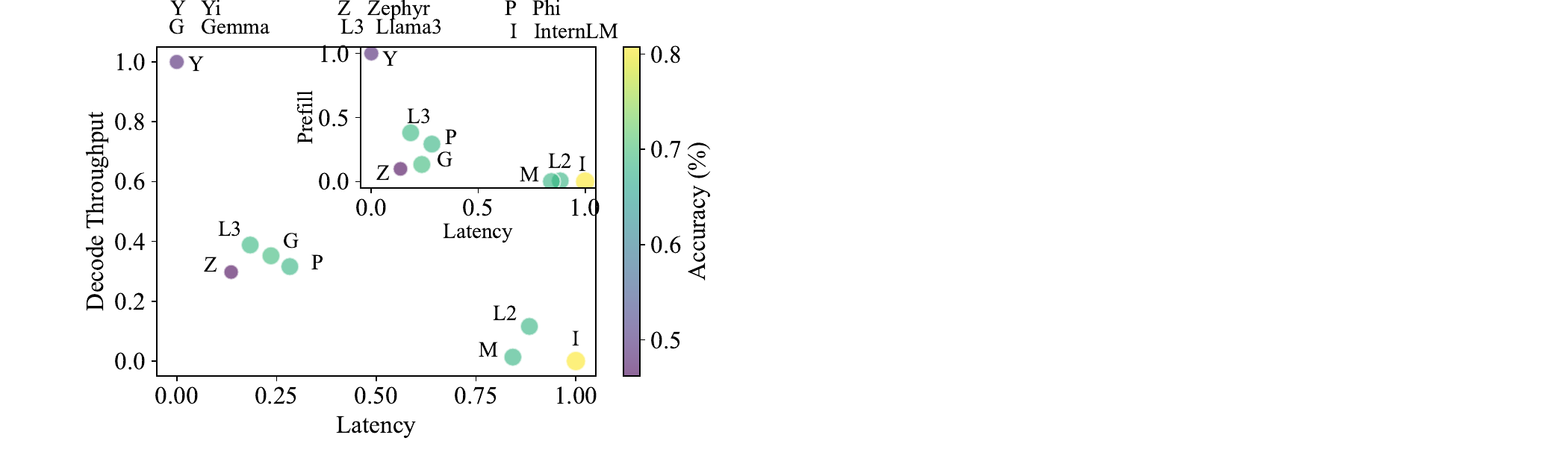}
\caption{Comparison of model performance across multiple metrics}
\label{fig:overall}
\vspace{-10pt}
\end{figure}

Figure.~\ref{fig:overall} reveals the performance characteristics of different models in terms of end-to-end latency, throughput, and accuracy. Large models, such as InternLM, Mistral, and Llama2, show significant trade-offs between latency and throughput, with InternLM having the highest latency and lowest throughput but achieving high accuracy. Medium-sized models, including Phi and Llama3, balance efficiency with capability. Llama3 demonstrates relatively lower latency and higher throughput compared to the large models, while Phi offers moderate efficiency. Lastly, small models, such as Yi, Zephyr, and Gemma, excel in latency and throughput. Yi leads with the lowest latency and highest throughput but sacrifices accuracy, while Gemma outperforms Yi in all metrics, though it lags behind Llama3.
\section{Conclusion}
This paper demonstrates the feasibility and limitations of deploying GenAI on edge devices, specifically targeting conversational AI applications. Using a Raspberry Pi-based edge computing testbed, we evaluated the performance of various LLMs across large, mid, and small model sizes. Our custom Kubernetes-based framework, built on APIs, supports modular integration and plug-and-play functionality, making it easy to add or swap models. By leveraging open-source tools, the testbed remains highly extensible and adaptable for ongoing and future experimentation. Evaluation results show that lightweight models such as Yi and Phi achieve sufficient throughput and latency for edge settings, presenting a viable solution for distributed intelligence without the need for specialized accelerators. 
In future work, we will to leverage the Testbed to develop a robust AI agent framework optimized for 6G network management. 

\bibliographystyle{unsrt}
\bibliography{Submitted}

\end{document}